\documentclass[acmtog, authorversion]{acmart}

\AtBeginDocument{%
  \providecommand\BibTeX{{%
    \normalfont B\kern-0.5em{\scshape i\kern-0.25em b}\kern-0.8em\TeX}}}

\usepackage{ifpdf,verbatim}

\ifpdf
\usepackage{framed,color}
\definecolor{shadecolor}{gray}{0.93}

\newenvironment{code}%
   {\snugshade\verbatim}%
   {\endverbatim\endsnugshade}

\else

\newenvironment{code}
    {\HCode{<div class='code'>}\verbatim}
    {\endverbatim\HCode{</div>}}

\fi

\acmBooktitle{}
\begin{document}

\title{Merly.jl: Web Framework in Julia}


\author{Josu\'e Acevedo Maldonado}
\affiliation{%
  \institution{@neomatrixcode}
  \city{Oaxaca}
  \country{M\'exico}}
\email{josuecevedo@gmail.com}
\orcid{0000-0002-4056-1895}

\renewcommand{\shortauthors}{Acevedo}

\begin{abstract}
  Merly.jl is a package for creating web applications in Julia. It presents features such as the creation of endpoints with function notation and with macro notation, handling of static files, use of path parameters, processing of data sent by a web client in the body in a personalized way, handling of CORS and compatibility with the use of middleware. It presents a familiar syntax with the rest of the most popular web frameworks without neglecting the execution performance. This manuscript mentions the operation and main features of Merly.jl
\end{abstract}



\maketitle

\section{Introduction}
At present, information systems have gone from running on a personal computer to working on the Internet, this change of environment has occurred mainly thanks to the evolution of protocols such as the Hypertext Transfer Protocol (HTTP) \cite{10.1145/1060745.1060746} , the creation of exchange formats of information such as XML\cite{10.1145/872757.872793} or JSON\cite{10.1145/2872427.2883029}, software architectures such as Microservices\cite{10.1145/3053600.3053653}  as well as Backend web development libraries. Such has been the impact that web development libraries exist in practically all programming languages, the best known being Flask, Sinatra or Express.

Julia is a programming language that has the main characteristic of using multiple dispatches and metaprogramming to provide flexibility in the syntax, such as execution efficiency on the machine \cite{bezanson2015julia}, offers the possibility of developing web applications, for the that requires the use of libraries that facilitate development and avoid reinventing the wheel. The most popular frameworks that Julia currently presents are: Mux.jl, Bukdu.jl and Dance.jl.

Merly.jl was born with the purpose of having a web framework that is easy for the developer to write; taking up syntax concepts from the most popular frameworks both from other languages and from Julia, as well as taking into account the performance of the code's execution.

\section{MERLY.JL}
\subsection{Handler operation}
Merly.jl creates an association using a path, an HTTP verb and an anonymous function AKA lambda \cite{10.5555/1717258}. When you request a URL with a certain HTTP verb in a web browser this function will run and display the result.

\begin{figure}[h]
  \centering
  \includegraphics[width=\linewidth]{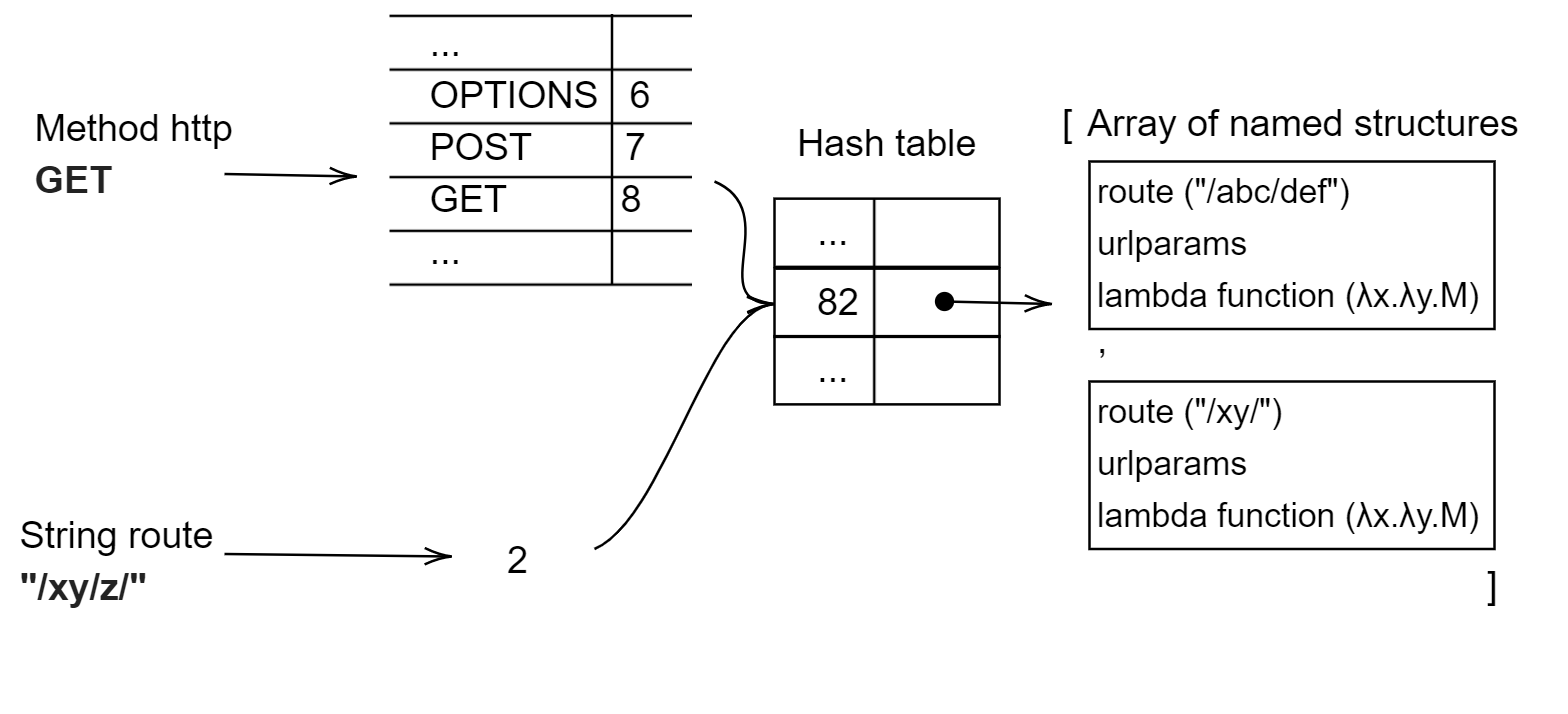}
  \caption{Create an endpoint in Merly.jl. A hash table is used due to the efficiency in the search process.}
  \Description{Diagram of creating an endpoint.}
\end{figure}

To obtain a good execution performance, the search process was prioritized so it was decided to use a hash table instead of a self-balancing binary tree or a single path array, which among other things present problems when the number of stored data is very high \cite{10.5555/280635, 10.5555/86560}. 

As the key of the hash table we have a number value, which is created from a value associated with each HTTP verb and which can be consulted in a table, which in practical terms is also a hash table, concatenated to the number of elements present in the path. This technique makes it possible to reduce the number of collisions when carrying out a search .

The value stored in the hash table will be an array of named structures, which will store the lambda function to be executed, such as the path and the parameters present in it. Being an array, this allows one or more routes to be added to it that have the same number of elements and the same HTTP verb. 

In such a way that, when performing the search, the hash table allows us to discard a large number of possible options with few computational calculations. On the resulting array of routes, we will perform a search with the match between the searched route and the previously stored route in parallel using the Iterators.filter method; which will finally obtain the function to be executed and whose result will be sent to the web client.

\begin{figure}[h]
  \centering
  \includegraphics[width=\linewidth]{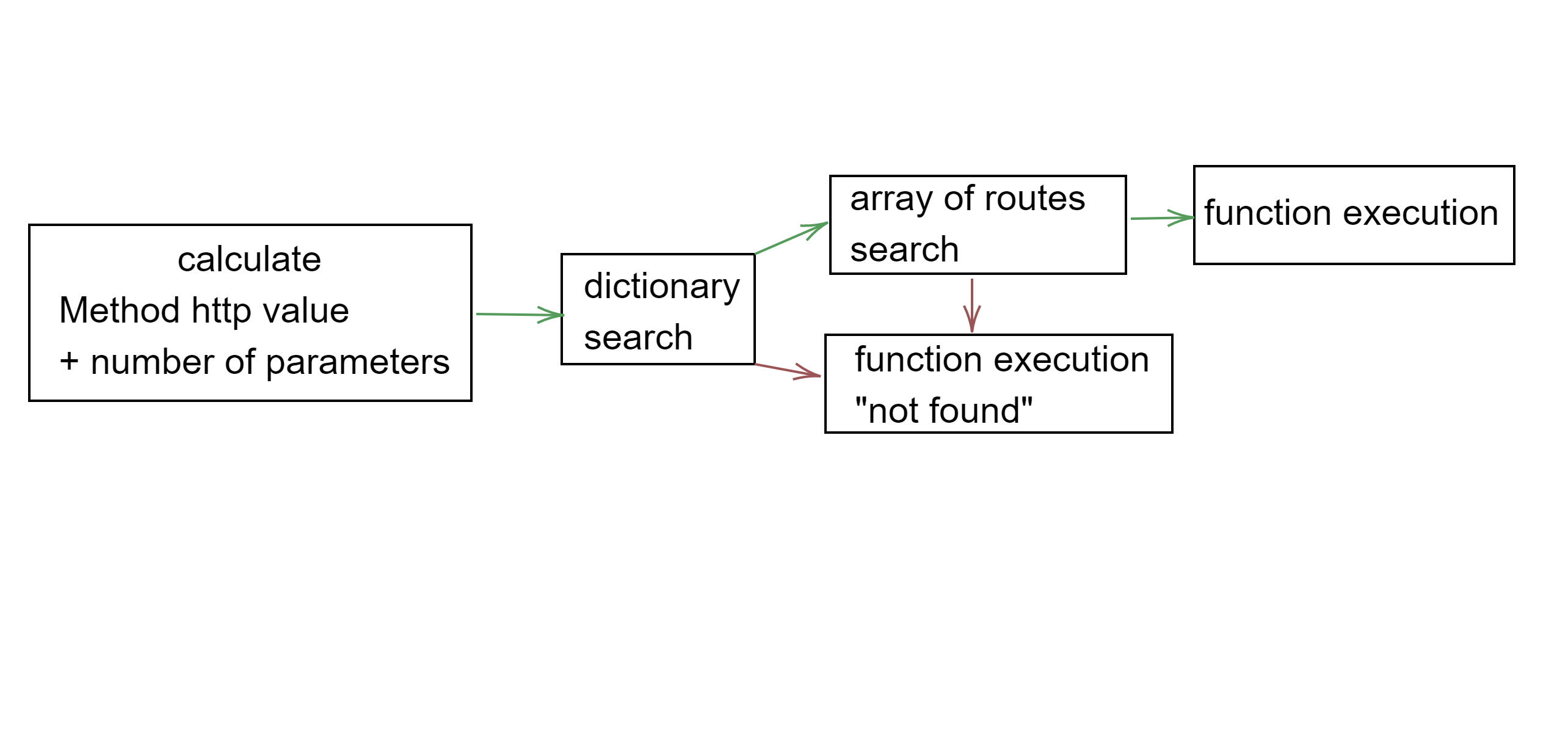}
  \caption{Search process and execution of an endpoint. The routes are obtained from the hash table and are analyzed in parallel to finally obtain the function to execute.}
  \Description{Execution diagram of an endpoint.}
\end{figure}

If a key is not found in the hash table that matches the client's request or, failing that, a path, a function stored with key 0 will be executed by default. This default function is known as “not found” and it is possible to replace it with other that executes our own custom code.

\subsection{Route paths}
Three elements are used to create route routes: an HTTP verb, a route, and an anonymous or lambda function. The easiest way to achieve this is by using functions, which respond to the name of each of the HTTP verbs. These functions receive as parameters a string that will define the path of the endpoint and an unnamed function, this lambda function always receives, when executed, a structure that will contain all the data sent by the client, called request, and a structure called HTTP ( defined in the HTTP .jl package) that will hold the HTTP response code, body, and headers.

\begin{code}

Get("/data", (request,HTTP)->begin
    HTTP.response(200, "test text Get")
end)

Post("/data", (request,HTTP)-> begin
    HTTP.response(200, "test text Post")
end)

Put("/data", (request,HTTP) -> begin
     HTTP.response(200, "test text Put")
end)
 
Delete("/data", (request,HTTP) -> begin
     HTTP.response(200, "test text Delete")
end)
 
Connect("/data", (request,HTTP) -> begin
     HTTP.response(200, "test text Connect")
end)
 
Trace("/data", (request,HTTP) -> begin
     HTTP.response(200, "test text Trace")
end)
 
Head("/data", (request,HTTP) -> begin
     HTTP.response(200, "test text head")
end)
 
Patch("/data", (request,HTTP) -> begin
     HTTP.response(200, "test text Patch")
end)
\end{code}

There is a shorter syntax with which route paths can be defined, that is, through Macros, which assign a tuple of arguments to an expression which is compiled directly. The macro @page implicitly assigns the GET method to a certain route and instead of defining a function, only its body is specified.

\begin{code}

@page "/" HTTP.Response(200,"Hello World!")
\end{code}

The @route macro, on the other hand, requires in addition to all the elements mentioned above the name of the HTTP verb, which allows the use of the operator "|" to define more than one verb to the same path and function body more easily.

\begin{code}

@route POST|PUT|DELETE "/route" begin
  println("query: ",request.query)
  println("body: ",request.body)

  HTTP.Response(200
        , HTTP.mkheaders(["Content-Type" => "text/plain"])
        , body="I did something!")
end
\end{code}

When macros receive only the body of the function as a parameter, it is necessary to define a lambda function using metaprogramming, so the context of this function will be different to the one of the program where the macro was used.

\subsection{Route parameters}

It is possible to define, in the path of the endpoint, segments or areas that will be occupied by data sent by the web client, for which the colon character ( : ) is used, thanks to which a name will be assigned to said parameter, this characteristic is compatible with both function notation and the two path definition syntaxes that use macros. If you want to use a regular expression to custom filter your path parameters, you can do it using a pair of parentheses and the backslash escape character ( \textbackslash ).

\begin{code}

@page "/hola/:usr" begin
  HTTP.Response(
  200
  ,string("<b>Hello ",request.params["usr"],"!</b>"))
end

@route GET "/get/:data1" begin
  HTTP.Response(200, string(u ,request.params["data1"]))
end

Get("/test1/:usr",
  (request, HTTP) -> begin
    HTTP.Response(
    200
    ,string("<b>test1 ",request.params["usr"],"!</b>")
    )
    end
)

@route GET "/regex/(\\w+\\d+)" begin
  
  return HTTP.Response(
  200
  ,string("datos ",request.params["2"])
  )
  
end
\end{code}

The data sent in the route will be accessible from the dictionary called params stored in the request structure, through the name defined when creating the route; by using a regular expression, the data can be accessed by referring to its position within the path.

\subsection{Variables}
Taking into account that the function associated with an endpoint will be executed or defined in a different context than the program where the route was established, variables can be passed to the context of the function to be executed, both in the function and Macro syntax. 

The data of these variables will be passed to the function either by reference or by value, depending on whether the data type is mutable or not, in the first case it will be passed by reference and in the second by value.

\begin{code}

u=1

@page "/get1" (;u=u) HTTP.Response(200
                         ,string("<b>Get1 ",u," !</b>"))

@route GET "/get/:data1" (;u=u) begin
  u = u +1
  HTTP.Response(200, string(u ,request.params["data1"]))
end
\end{code}

\subsection{Server}
To process the data from the web client, it is necessary that there is a process that listens for requests. To configure the execution options of this process such as the host, the port and the ssl keys, named variables are used. These are received by the function called \verb|start|. 

The host parameter receives a text string with the ip address, which can be version 4 or version 6. To manage the configuration of the cryptographic keys to use an encrypted HTTP connection, the MbedTLS package is used, specifically the \verb|SSLConfig| function which will process the files of both the certificate and the key.

\begin{code}

using MbedTLS
  
start( 
  host = "127.0.0.1"
, port = 8086
, sslconfig= MbedTLS.SSLConfig("localhost.crt"
                                   ,"localhost.key") )
\end{code}

\subsection{Headers}
The values of the headers can be specified at the time of returning the HTTP structure using the \verb|mkheaders| function. However, if you want to constantly return one or more specific headers you can rely on the \verb|headersalways| function which receives an array of string 2-tuples. This way code redundancy is avoided.

\begin{code}

headersalways([
  "X-PINGOTHER" => "pingpong"
, "Y-PINGOTHER" => "text"])
\end{code}

 If you ever need to use CORS, Merly has you covered, there is a dedicated function to configure the CORS headers in addition to enabling  the necessary OPTION routes for the verified CORS requests to work without too much complexity.

\begin{code}

useCORS(
  AllowOrigins = "*"
, AllowHeaders = "Origin, Content-Type, Accept"
, AllowMethods = "GET,POST,PUT,DELETE"
, MaxAge = "178000")
\end{code}

The \verb|useCORS| function receives a set of named variables which are:

\begin{itemize}
\item AllowOrigins: Specifies a URI that can access the resource.
\item AllowHeaders: Indicates which HTTP header can be used when requesting the resource.
\item AllowMethods: Specifies the method or methods allowed when a resource is allocated.
\item MaxAge: This header indicates for how long the results of the verified request can be captured, and therefore it is no longer necessary to carry out the process again.
\end{itemize}

\subsection{Static files}

If we need to associate a url with a file stored on the computer, allowing the access to its content through the http protocol we can use the \verb|File| and \verb|webserverfiles| functions.

\verb|File| is a function that receives a text string that includes the name of a file with its perspective extension and returns the content of that file in a text string format. By default, this file must be in the same location where the program is executed, considered as the root or current directory, otherwise the relative path from this current location must be included.

\begin{code}

  File("index.html")
\end{code}

The \verb|webserverfiles| function will take all the existing files in the root or current directory and expose them, although there is a parameter that allows controlling this behavior; if the \verb|webserverfiles| function receives a text string equal to ( * ) it will expose all files without exception, but if you want to discriminate files by extension, then a text string with the extension or extensions separated by pipe will be passed to the function ( | ) of the files to expose.

\begin{code}

  webserverfiles("*")
\end{code}

There is the possibility of modifying the directory considered as root, with the \verb|webserverpath| function that receives a text string with the name of the folder or its full path, in such a way that when executing either the \verb|File| or \verb|webserverfiles| function, you will be working with the existing files in the new location.

\begin{code}

   webserverpath("folder") 
\end{code}

\subsection{Body parser}
The data sent by the client in the body field can be of any type, by default Merly.jl processes the data as text strings, so within the function this data can be found in the body field of the structure request.

If at any time, you want to work with julia's native dictionaries to process the data, the conversion process would have to be done manually each time is required. So an easy way to make Merly.jl deliver this data already processed is with a custom function that receives a text string by default and returns the data already processed, with an existing library or with its own algorithm.

\begin{code}

function tojson(data::String)
   return JSON.parse(data)
end

formats["application/json"] = tojson
\end{code}

Subsequently, the name of this function is stored in a dictionary called \verb|formats|, provided by the package; the key associated with this function being a text string with a specific Content-Type. There is the possibility of processing each format or data type sent by the client in a different way, for example, XML, JSON, etc.

\begin{code}

  Post("/data", (request,HTTP)-> begin

  HTTP.Response(200
        , HTTP.mkheaders(["Content-Type" => "text/plain"])
        , body=string("I did something! "
                                 , request.body["query"]))

end)
\end{code}

\subsection{Middleware}
At the moment, this feature is only available in function notation. However, it extends the functionalities of the functions to be executed by allowing them to concatenate with other functions that perform tasks either before or after the main function.

\begin{code}

function authenticate(request, HTTP)
  isAuthenticated = false

  if (request.params["status"] === "authenticated")
    isAuthenticated = true
  end

  return request, HTTP, isAuthenticated
end

Get("/verify/:status",

(result(;middleware=authenticate) = (request, HTTP)->
begin

  myfunction = (request, HTTP, isAuthenticated)-> begin

    if (isAuthenticated == false )
          return  HTTP.Response(
                  403
                  ,"Unauthenticated. Please signup!"
                )
    end
    return  HTTP.Response(200,string("<b>verify !</b>"))
    
    end

  return myfunction(middleware(request,HTTP)...)

end
)()
)
\end{code}

\subsection{Future work}
In future versions of Merly.jl it is planned to add the possibility of working with websockets, make the middleware writing easier and optimize memory use when creating routes.

\subsection{Quality Control}
The Project presents a set of tests hosted in the runtests.jl file, which tests most of the Merly.jl features and runs automatically on the TravisCI platform on Windows, Linux and Mac OSX operating systems and on the Julia versions 1.5, 1.6 and 1.7. Also, the amount of code evaluated by the tests is monitored on the AppVoyer platform.

Locally, these same tests can be executed with the Pkg.test ("Merly") command, so any user can verify that the package works correctly on their computer. If there is a problem with the operation of the package or want to propose new features, users can go to the project's issues page on Github and open a new one.

\section{AVAILABILITY}
\subsection{Operating System}
Merly works on Windows 7+, Mac OSX, Linux and FreeBSD

\subsection{Programming language}
Julia v1.5+

\subsection{Dependencies}
The HTTP.jl package is used to manage communication through the HTTP protocol, as well as MbedTLS.jl to manage the cryptographic keys that allow the use of the HTTPS protocol.

\subsection{List of contributors}
\begin{itemize}
\item Josu\'e Acevedo Maldonado, lead developer of Merly.jl
\item Phelipe Wesley, contributed by rewriting a large part of the package to improve its compatibility with the 1.4 version of Julia
\end{itemize}

\subsection{Software location}
\begin{itemize}
\item File: Zenodo

Name: Merly.jl

Persistent identifier: DOI: \href{https://doi.org/10.5281/zenodo.4546005}{10.5281/zenodo.4546005}

License: MIT

Publisher: Acevedo Maldonado Josu\'e

Version published: v1.0.0

\item Code repository: GitHub

Name: neomatrixcode/Merly.jl

Persistent identifier: \url{https://github.com/neomatrixcode/Merly.jl}

License: MIT

Date published: 07/02/2021

Version published: v1.0.0

Documentation Language: English

Programming Language: Julia
\end{itemize}

\section{CONCLUDING REMARKS }
Julia is a relatively recent programming language which has had the greatest boom so far in the field of data science and artificial intelligence. However, these are not the only areas where the language can address thanks to its features such as ease of use and execution speed. Julians who are looking for a package that allow them to expose resources on the web in a simple and efficient way with periodic updates of features will find in Merly.jl an important ally to achieve their objectives.

\begin{acks}
A thank you to the Julia community for their contributions to the package and their feedback on the features to add or modify in this package, also to my friend Vásquez Martínez Agustín (@maldad) for his help in reviewing the first versions of this document.
\end{acks}

\bibliographystyle{ACM-Reference-Format}
\bibliography{sample-base}

\appendix

\section{Online Resources}

On the Merly.jl website you can find \href{https://merly.vercel.app/#examples}{practical examples} to use the framework to create a web application and deploy it to different cloud providers, as well as Docker and Docker-compose.

\end{document}